# Are We Ready For Learned Cardinality Estimation?

Xiaoying Wang[†*], Changbo Qu[†*], Weiyuan Wu[†*], Jiannan Wang[†], Qingqing Zhou[◇]
Simon Fraser University[†]　　　　Tencent Inc.[◇]
{xiaoying_wang, changboq, youngw, jnwang}@sfu.ca　　　hewanzhou@tencent.com

## ABSTRACT

Cardinality estimation is a fundamental but long unresolved problem in query optimization. Recently, multiple papers from different research groups consistently report that learned models have the potential to replace existing cardinality estimators. In this paper, we ask a forward-thinking question: *Are we ready to deploy these learned cardinality models in production?* Our study consists of three main parts. Firstly, we focus on the static environment (i.e., no data updates) and compare five new learned methods with nine traditional methods on four real-world datasets under a unified workload setting. The results show that learned models are indeed more accurate than traditional methods, but they often suffer from high training and inference costs. Secondly, we explore whether these learned models are ready for dynamic environments (i.e., frequent data updates). We find that they cannot catch up with fast data updates and return large errors for different reasons. For less frequent updates, they can perform better but there is no clear winner among themselves. Thirdly, we take a deeper look into learned models and explore when they may go wrong. Our results show that the performance of learned methods can be greatly affected by the changes in correlation, skewness, or domain size. More importantly, their behaviors are much harder to interpret and often unpredictable. Based on these findings, we identify two promising research directions (control the cost of learned models and make learned models trustworthy) and suggest a number of research opportunities. We hope that our study can guide researchers and practitioners to work together to eventually push learned cardinality estimators into real database systems.

## 1 INTRODUCTION

The rise of "ML for DB" has sparked a large body of exciting research studies exploring how to replace existing database components with learned models [32, 37, 39, 68, 84, 98]. Impressive results have been repeatedly reported from these papers, which suggest that "ML for DB" is a promising research area for the database community to explore. To maximize the impact of this research area, one natural question that we should keep asking ourselves is: *Are we ready to deploy these learned models in production?*

In this paper, we seek to answer this question for cardinality estimation. In particular, we focus on *single-table cardinality estimation*, a fundamental and long standing problem in query optimization [18, 95]. It is the task of estimating the number of tuples of a table that satisfy the query predicates. Database systems use a query optimizer to choose an execution plan with the estimated minimum cost. The performance of a query optimizer largely depends on the quality of cardinality estimation. A query plan based on a wrongly estimated cardinality can be orders of magnitude slower than the best plan [42].

Multiple recent papers [18, 28, 30, 34, 95] have shown that learned models can greatly improve the cardinality estimation accuracy compared with traditional methods. However, their experiments have a number of limitations (see Section 2.5 for more detailed discussion). Firstly, they do not include all the learned methods in their evaluation. Secondly, they do not use the same datasets and workload. Thirdly, they do not extensively test how well learned methods perform in dynamic environments (e.g., by varying update rate). Lastly, they mainly focus on when learned methods will go right rather than when they may go wrong.

We overcome these limitations and conduct comprehensive experiments and analyses. The paper makes four contributions:

**Are Learned Methods Ready For Static Environments?** We propose a unified workload generator and collect four real-world benchmark datasets. We compare five new learned methods with nine traditional methods using the same datasets and workload in static environments (i.e., no data updates). The results on accuracy are quite promising. In terms of training/inference time, there is only one method [18] that can achieve similar performance with existing DBMSs. The other learned methods typically require $10 - 1000\times$ more time in training and inference. Moreover, all learned methods have an extra cost for hyper-parameter tuning.

**Are Learned Methods Ready For Dynamic Environments?** We explore how each learned method performs by varying update rate on four real-world datasets. The results show that learned methods fail to catch up with fast data updates and tend to return large error for various reasons (e.g., the stale model processes too many queries, the update period is not long enough to get a good updated model). When data updates are less frequent, learned methods can perform better but there is no clear winner among themselves. We further explore the update time vs. accuracy trade-off, and investigate how much GPU can help learned methods in dynamic environments.

**When Do Learned Methods Go Wrong?** We vary correlation, skewness, and domain size, respectively, on a synthetic dataset, and try to understand when learned methods may go wrong. We find that all learned methods tend to output larger error on more correlated data, but they react differently w.r.t. skewness and domain size. Due to the use of black-box models, their wrong behaviors are very hard to interpret. We further investigate whether their behaviors follow some simple and intuitive logical rules. Unfortunately, most of them violate these rules. We discuss four issues related to deploying (black-box and illogical) learned models in production.

[*] The first three authors contributed equally to this research.

Table 1: Taxonomy of New Learned Cardinality Estimators.

|  | Methodology | Input | Model |
|---|---|---|---|
| **MSCN** [34] | Regression | Query+Data | Neural Network |
| **LW-XGB** [18] | Regression | Query+Data | Gradient Boosted Tree |
| **LW-NN** [18] | Regression | Query+Data | Neural Network |
| **DQM-Q** [28] | Regression | Query | Neural Network |
| **Naru** [95] | Joint Distribution | Data | Autoregressive Model |
| **DeepDB** [30] | Joint Distribution | Data | Sum Product Network |
| **DQM-D** [28] | Joint Distribution | Data | Autoregressive Model |

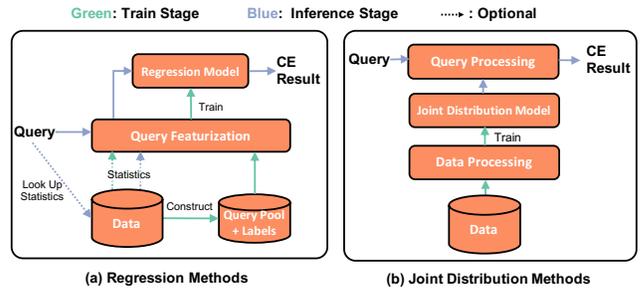

Figure 1: Workflow of Learned Methods.

**Research Opportunities.** We identify two future research directions: i) control the cost of learned methods and ii) make learned methods trustworthy, and suggest a number of promising research opportunities. We publish our code and datasets on GitHub[1] to facilitate future research studies. We hope our work can attract more research efforts in these directions and eventually overcome the barriers of deploying learned estimators in production.

The rest of the paper is organized as follows: We present a survey on learned cardinality estimation in Section 2 and describe the general experimental setup in Section 3. We explore whether learned methods are ready for static environments in Section 4 and for dynamic environments in Section 5, and examine when learned methods go wrong in Section 6. Future research opportunities are discussed in Section 7. Multi-table scenario are discussed in Section 8 and related works are reviewed in Section 9. Finally, we present our conclusions in Section 10.

## 2 LEARNED CARDINALITY ESTIMATION

In this section, we first formulate the cardinality estimation (CE) problem, then put new learned methods into a taxonomy and present how each method works, and finally discuss the limitations of existing evaluation on learned methods.

### 2.1 Problem Statement

Consider a relation $R$ with $n$ attributes $\{A_1, \ldots, A_n\}$ and a query over $R$ with a conjunctive of $d$ predicates:

SELECT COUNT(*) FROM R
WHERE $\theta_1$ AND $\cdots$ and $\theta_d$,

where $\theta_i$ ($i \in [1, d]$) can be an *equality predicate* like $A = a$, an *open range predicate* like $A \leq a$, or a *close range predicate* like $a \leq A \leq b$. The goal of CE is to estimate the answer to this query, i.e., the number of tuples in $R$ that satisfy the query predicates. An equivalent problem is called *selectivity estimation*, which computes the *percentage* of tuples that satisfy the query predicates.

### 2.2 Taxonomy

The idea of using ML for CE is not new (see Section 9 for more related work). The novelty of recent learned methods is to adopt more advanced ML models, such as deep neural networks [18, 28, 34], gradient boosted trees [18], sum-product networks [30], and deep autoregressive models [28, 95]. We call these methods "new learned methods" or "learned methods" if the context is clear. In contrast, we refer to "traditional methods" as the methods based on histogram or classic ML models like KDE and Bayesian Network.

Table 1 shows a taxonomy of new learned methods[2]. Based on the methodology, we split them into two groups - *Regression* and *Joint Distribution* methods. *Regression* methods (a.k.a *query-driven* methods) model CE as a regression problem and aim to build a mapping between queries and the CE results via feature vectors, i.e., $query \rightarrow feature\_vector \rightarrow CE\_result$. *Joint Distribution* methods (a.k.a *data-driven* methods) model CE as a joint probability distribution estimation problem and aim to construct the joint distribution from the table, i.e., $P(A_1, A_2, \cdots, A_n)$, then estimate the cardinality. The Input column represents what is the input to construct each model. Regression methods all require queries as input while joint distribution methods only depend on data. The Model column indicates which type of model is used correspondingly. We will introduce these methods in the following.

### 2.3 Methodology 1: Regression

**Workflow.** Figure 1(a) depicts the workflow of regression methods. In the *training stage*, it first constructs a query pool and gets the label (CE result) of each query. Then, it goes through the query featurization module, which converts each query to a feature vector. The feature vector does not only contain query information but also optionally include some statistics (like a small sample) from the data. Finally, a regression model is trained on a set of ⟨feature vector, label⟩ pairs. In the *inference stage*, given a query, it converts the query to a feature vector using the same process as the training stage, and applies the regression model to the feature vector to get the CE result. To handle *data updates*, regression methods need to update the query pool and labels, generate new feature vectors, and update the regression model.

There are four regression methods: MSCN, LW-XGB, LW-NN, and DQM-Q. One common design choice in them is the usage of log-transformation on the selectivity label since the selectivity often follows a skewed distribution and log-transformation is commonly used to handle this issue [19]. These works vary from many perspectives, such as their input information, query featurization, and model architecture.

**MSCN [34]** introduces a specialized deep neural network model termed multi-set convolutional network (MSCN). MSCN can support join cardinality estimation. It represents a query as a feature vector which contains three modules (i.e., table, join, and predicate modules). Each module is a two-layer neural network and different module outputs are concatenated and fed into a final output network, which is also a two-layer neural network. MSCN enriches the training data with a materialized sample. A predicate will be evaluated on a sample, and a bitmap, where each bit indicates whether a tuple in the sample satisfies the predicate or not, will be added to the feature vector. This enrichment has been proved to make obvious positive impact on the model performance [34, 95].

---

[1] https://github.com/sfu-db/AreCELearnedYet

[2] Naru, DeepDB and MSCN are named by their authors. For convenience of discussion, we give others the following short names. Lightweight Gradient Boosting Tree (LW-XGB) and Lightweight Neural Network (LW-NN) are two models from [18]. From [28], two complementary methods are proposed, Data&Query Model - Data (DQM-D) and Data&Query Model - Query (DQM-Q).

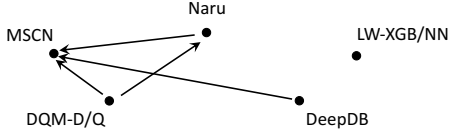

Figure 2: Comparison results available in existing studies.

**LW-XGB/NN [18]** introduces a lightweight selectivity estimation method. Its feature vector consists of two parts: range features + CE features. The range features represent a set of range predicates: $\langle a_1, b_1, a_2, b_2, \cdots, a_n, b_n \rangle$. The CE features represent heuristic estimators (e.g., the one that assumes all columns are independent). Note that the CE features can be cheaply derived from the statistics available in the database system. LW-NN (LW-XGB) train a neural network (gradient boost tree) model using the generated features. Unlike MSCN which minimizes the mean q-error, they minimize the mean square error (MSE) of the log-transformed label, which equals to minimizing the geometric mean of q-error with more weights on larger errors and also can be computed efficiently.

**DQM-Q [28]** proposes a different featurization approach. It uses one-hot encoding to encode categorical columns and treats numerical attributes as categorical attributes by automatic discretization [15]. DQM-Q trains a neural network model. When a real-world query workload is available, DQM-Q is able to augment the training set and train the model with the augmented set.

## 2.4 Methodology 2: Joint Distribution

**Workflow.** Figure 1(b) depicts the workflow of joint distribution methods. In the *training stage*, it transforms the data into a format ready for training a joint distribution model. In the *inference stage*, given a query, it generates one or multiple requests to the model and combine the model inference results into the final CE result. To handle *data updates*, joint distribution methods need to update or retrain the joint distribution model.

There are three joint distribution methods: Naru, DeepDB, and DQM-D. Compared to traditional methods like histogram and sampling, these new methods adopt more complex models to further capture additional information in the data, such as fine-grained correlation or conditional probability between columns.

**Autoregressive Model.** Naru [95] and DQM-D [28] propose similar ideas. They factorize the joint distribution into conditional distributions using the product rule: $P(A_1, A_2, ..., A_n) = P(A_1)P(A_2|A_1) \cdots P(A_n|A_1, ..., A_{n-1})$. They adopt the state-of-the-art deep autoregressive models such as MADE [23] and Transformer [89] to approximate the joint distribution.

The joint distribution can directly return results to point queries. To support range queries, they adopt a sampling based method, which runs importance sampling in an adaptive fashion. Specifically, Naru uses a novel approximation technique named progressive sampling, which samples values column by column according to each internal output of conditional probability distribution. DQM-D adopts an algorithm [44] originally designed for Monte-Carlo multi-dimensional integration, which conducts multiple stages of sampling. At each stage, it selects sample points in proportion to the contribution they make to the query cardinality according to the result from the previous stage.

**Sum-Product Network.** DeepDB [30] builds Sum-Product Networks (SPNs) [72] to capture the joint distribution. The key idea is to recursively split the table into different clusters of rows (creating a sum node to combine them) or clusters of columns (assuming different column clusters are independent and creating a product node to combine them). KMeans is used to cluster rows and Randomized Dependency Coefficients [50] is used to identify independent columns. Leaf nodes in an SPN represent a single attribute distribution, which can be approximated by histograms for discrete attributes or piecewise linear functions for continuous attributes.

## 2.5 Limitations of Existing Experiments

As pointed in the Introduction, existing experimental studies have a number of limitations. We provide more detail in this section.

Firstly, many new learned methods have not been compared with each other directly. Figure 2 visualizes the available comparison results using a directed graph. Each node represents a method, and if method A has compared with method B in A's paper, we draw a directed edge from A to B. Since many methods were proposed in the same year or very close period, the graph is quite sparse and misses over half of the edges. For example, LW-XGB/NN is one of the best regression methods, but it has no edge with any other method. DeepDB and Naru are two state-of-the-art joint distribution methods, but there is no edge between them.

Secondly, there is no standard about which datasets to use and how to generate workloads. Other than the IMDB dataset (adopted by MSCN and DeepDB), none of the datasets adopted in one work appear in another work. As for workloads, these works generate synthetic queries differently. Table 2 compares their generated workloads. For join queries in the JOB-light benchmark (used in MSCN and DeepDB), we report their properties related to single table. $|D|$ denotes the number of columns in the dataset and OOD (out-of-domain) means that the predicates of a query are generated independently. Such queries often lead to zero cardinality.

Thirdly, existing works are mostly focused on the static environment (i.e., no data update setting). However, dynamic environments are also common in practice. Some papers have explored how their method performs when the data updates, but the way that they update the data varies. As a result, the performance numbers cannot be used to compare between methods. Furthermore, existing studies have not extensively explored the trade-off between accuracy and updating time. For example, Naru is a more accurate method but requires longer time to update the model. It is unclear whether it can still give good accuracy for high update rates.

## 3 EXPERIMENTAL SETUP

Our study evaluates learned cardinality estimators under different settings. We describe the general setup used in all of our experiments in this section.

**Evaluation Metric.** We use q-error as our accuracy metric to measure the quality of the estimation result. Q-error is a symmetric metric which computes the factor by which an estimate differs from the actual cardinality: $error = \frac{max(est(q), act(q))}{min(est(q), act(q))}$. For example, if a query's actual cardinality is 10 and estimated cardinality is 100, then $error = \frac{max(100, 10)}{min(100, 10)} = 10$.

Table 2: Workload used in existing experimental studies.

|  | Predicate Number | Operator Equal | Operator Range | Consider OOD |
|---|---|---|---|---|
| **MSCN** | $0 \sim |D|$ | ✓ | ✓ | ✗ |
| **LW-XGB/NN** | $2 \sim |D|$ | ✗ | close range | ✓ |
| **Naru** | $5 \sim 11$ | ✓ | open range | ✓ |
| **DeepDB** | $1 \sim 5$ | ✓ | ✓ | ✗ |
| **DQM-D/Q** | $1 \sim |D|$ | ✓ | ✗ | ✓ |
| **Our Workload** | $1 \sim |D|$ | ✓ | ✓ | ✓ |

Table 3: Dataset characteristics. "Cols/Cat" means the number of columns and categorical columns; "Domain" is the product of the number of distinct values for each column.

| Dataset | Size(MB) | Rows | Cols/Cat | Domain |
| --- | --- | --- | --- | --- |
| Census [16] | 4.8 | 49K | 13/8 | $10^{16}$ |
| Forest [16] | 44.3 | 581K | 10/0 | $10^{27}$ |
| Power [16] | 110.8 | 2.1M | 7/0 | $10^{17}$ |
| DMV [62] | 972.8 | 11.6M | 11/10 | $10^{15}$ |

Q-error is the metric adopted by all learned methods [18, 28, 30, 34, 95]. It measures the relative error, which can penalize large and small results to the same extent. Furthermore, it has been proved to be directly related to the plan quality in query optimization [59].

**Learned Methods & Implementation.** As shown in Table 1, there are five recently published papers on learned methods: Naru [95], MSCN [34], LW-XGB/NN [18], DeepDB [30], and DQM [28]. We exclude DQM from our study since its data driven model has a similar performance with Naru and its query driven model does not support our workload (confirmed with DQM's authors).

For Naru[3] and DeepDB[4], we adopt the implementation released by the authors with minor modifications in order to support our experiments. We choose ResMADE as basic autoregressive building block for Naru because it is both efficient and accurate. For MSCN, since the original model supports join query, it needs extra input features to indicate different joins and predicates on different tables. To ensure a fair comparison on single table cardinality estimation, we modify the original code[5] by only keeping features represent predicates and qualifying samples. We implement both neural network (LW-NN, on PyTorch [67]) and gradient boosted tree (LW-XGB, on XGBoost [10]) approach for LW-XGB/NN according to the description in its original paper [18], and use Postgres's estimation result on single column to compute the CE features.

All the code including dataset manipulation, workload generation and estimator evaluation are released[6].

**Hardware and Platform.** We perform our experiments on a server with 16 Intel Xeon E7-4830 v4 CPUs (2.00GHz). For the neural network models (Naru, MSCN, LW-NN), we run them not only on CPU but also on a NVIDIA Tesla P100 GPU to gain more insights under different settings.

## 4 ARE LEARNED METHODS READY FOR STATIC ENVIRONMENTS?

Are learned estimators more accurate than traditional methods in static environment? What is the cost for the high accuracy? In this section, we first compare the accuracy of learned methods with traditional methods, and then measure their training and inference time in order to see whether they are ready for production.

### 4.1 Setup

**Dataset.** We use four real-world datasets with various characteristics (Table 3). We choose these datasets because first, the size of these datasets are in different magnitudes and the ratio between categorical and numerical columns varies; second, each dataset has been used in the evaluation of at least one prior work in this field.

**Workload.** We propose a unified workload generator. The goal of our workload generator is to be able to cover all the workload settings used in existing learned methods (see Table 2). We apply the same generator setting on all datasets in the same experiment.

---
[3]https://github.com/naru-project/naru
[4]https://github.com/DataManagementLab/deepdb-public
[5]https://github.com/andreaskipf/learnedcardinalities
[6]https://github.com/sfu-db/AreCELearnedYet

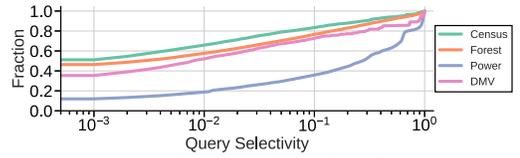

Figure 3: Distribution of workload selectivity.

Intuitively, a query with $d$ predicates can be thought of as a hyper-rectangle in a $d$-dimensional space. A hyper-rectangle is controlled by its center and width. Correspondingly, a query is controlled by its *query center* and *range width*. For example, consider a query with $d = 2$ predicates:

```
SELECT COUNT(*) FROM R
WHERE 0 ≤ A₁ ≤ 20 AND 20 ≤ A₂ ≤ 100
```

Its query center is $(\frac{20-0}{2}, \frac{100-20}{2}) = (10, 40)$ and its range width is $(20 - 0, 100 - 20) = (20, 80)$.

There are two ways to generate query centers. For ease of illustration, suppose that we want to generate a query center for columns $A_1, A_2$. The first way (①) is to randomly select a tuple $t$ from the table. Let $t[A_1], t[A_2]$ denote the attribute values of the tuple on $A_1$ and $A_2$. Then, we set the query center to $(t[A_1], t[A_2])$. The second way (②) is to independently draw a random value $c_1$ and $c_2$ from the domain of $A_1$ and $A_2$, respectively, and set the query center to $(c_1, c_2)$. ② is called out-of-domain (OOD in Table 2), which aims to test the robustness of learned estimators more comprehensively from the entire joint domain.

There are two ways to generate range widths. Let the domain for $A_i$ be $[\min_i, \max_i]$ and the domain size be $\text{size}_i = \max_i - \min_i$. The first way (❶) is to uniformly select a value $w_i$ from $[0, \text{size}_i]$. The second way (❷) is to select a value from an exponential distribution with a parameter $\lambda_i$ (we set $\lambda = 10/\text{size}_i$ by default). Note that if $A_i$ is a categorical column, we will only generate an equality predicate for it, thus the width is set to zero in this case. If a range on one side is larger than $\max_i$ or smaller than $\min_i$, then it becomes an open range query. Thus, our workload contains both open and close range queries.

Our workload generator covers all the above settings (①, ②, ❶, ❷). To generate a query, we first uniformly select a number $d$ from 1 to $|D|$ and randomly sample $d$ distinct columns to place the predicates. The query center is generated from ① and ② with a probability of 90% and 10%, respectively, and the range width is generated from ❶ and ❷ in equal proportions. The reason that we do not use an equal probability for the query center is that OOD is typically less common than the other way in real workloads. Figure 3 shows the selectivity distribution of generated workloads on different datasets, which results in a broad spectrum.

**Hyper-parameter Tuning.** We describe hyper-parameter tuning for each model. More details can be found in our Github repository.

For neural network methods (Naru, MSCN, LW-NN), we control the model size within 1.5% of the data size for each dataset. For each method, we select four model architectures with different numbers of layers, hidden units, embedding size, etc. and train each model in different batch size and learning rate in accordance with the original papers. Since MSCN and LW-NN are query-driven methods, we select 10K queries as a validation set to determine which hyper-parameters are better. Since Naru is a data-driven method (i.e., no query as input), we use training loss to find optimal hyper-parameters.

For LW-XGB, we vary the number of trees (16, 32, 64...) as in [18]. Since LW-XGB is a query-driven method, similar to MSCN and LW-NN, we select 10K validation queries for it.

For DeepDB, we do a grid search on RDC threshold and minimum instance slice and only keep the models within the size budget (i.e., 1.5% of the data size). An interesting finding is that DeepDB does not output the training loss like Naru during construction, thus queries are needed for hyper-parameter tuning. However, DeepDB is designed to be a data-driven method, which is not supposed to use queries. To ensure a fair comparison with other methods, we select a very small number of validation queries (i.e., 100 queries) for DeepDB to do hyper-parameter tuning.

To ensure a fair comparison, we use 100K queries to train all the query-driven methods (MSCN, LW-XGB/NN).

**Traditional Techniques.** We compare with a variety of traditional techniques, which are either used by real database systems or reported to achieve the state-of-the-art performance recently. The methods we chose can represent a wide range of solutions.

- *Postgres, MySQL and DBMS-A* are used to represent the performance of real database systems. We use PostgreSQL 11.5 and 8.0.21 MySQL Community Server, and DBMS-A is a leading commercial database system. They estimate cardinality rapidly with simple statistics and assumptions. In order to let them achieve their best accuracy level, we set the number of histogram buckets to the upper limit (10,000 for Postgres, 1024 for MySQL). For DBMS-A, we create several multi-column statistics in order to cover all columns with histograms. Note that even with the maximum number of buckets, size of these statistics is much smaller than our size budget, and result in less memory consumption than other traditional and learned methods in our experiment.
- *Sample-A, Sample-B* exhibit estimators adopt sampling. Sample-A uses a uniform random sample, which is well known that it would result in large error when no tuple in the sample satisfies all the predicates. Therefore we also include Sample-B, which assumes independence between each predicate in zero-tuple cases. We sample 1.5% tuples from each dataset for both methods.
- *MHIST* [73] builds a multi-dimensional histogram on the entire dataset. We choose Maxdiff as the partition constraint with Value and Area being the sort and source parameter since it is the most accurate choice according to [74]. We run the MHIST-2 algorithm iteratively until it reaches to 1.5% of the data size.
- *QuickSel* [66] represents query-driven multi-dimensional synopsis approaches' performance. It models the data distribution with uniform mixture model by leveraging query feedback. We choose QuickSel because it shows better accuracy than query-driven histograms including STHoles [6] and ISOMER [81] in [66]. We use 10K queries to train the model.
- *Bayes* [13] shows the estimation results of probabilistic graphical model approaches [14, 24, 88]. We adopt the same implementation in [95], which uses progressive sampling to estimate range queries and shows a very promising accuracy.
- *KDE-FB* [29] represents the performance of modeling data distribution with kernel density models. It improves naive KDE by optimizing the bandwidth with query feedback. We sample 1.5% tuples from each dataset (max to 150K) and use 1K queries to train the model.

### 4.2 Are Learned Methods More Accurate?

We first want to understand the learned methods' effort on accuracy improvement, comparing with traditional methods. We test all the methods using 10K queries on each dataset. Table 4 shows the q-error comparison result. Bold values in the "Traditional Methods" section denotes the minimum q-error that traditional methods can reach, while in the "Learned Methods" section it highlights the learned methods that can achieve a smaller (or equal) q-error than the best traditional method. The last row summaries the comparison by using "win" to denote learned methods beating traditional methods, and "lose" means the opposite.

Overall, learned methods are more accurate than traditional methods in almost all the scenarios. The best learned method can beat the best traditional method up to 14× on max q-error. The improvement over the three real database systems is particularly impressive. For example, they achieve 28×, 51×, 938×, and 1758× better max q-error on Census, Forest, Power and DMV, respectively. Even in the only exception that learned methods lose (50th on Forest), they can still achieve very similar performance to the best traditional result.

To see how the predicates affect the accuracy, we group the test queries based on the number of predicates and plot the q-error distribution of each group on Census dataset. Figure 4 shows the result. "Best Learned" (or "Best Traditional") represents the minimum q-error that learned (or traditional) methods can achieve on each percentile (max, 75th, median, 25th, min) value in the boxplot. We can see that the performance degrades when the number of predicates increases for both methods. It is because queries with more predicates tend to result in lower selectivity and the correlation between attributes tend to be more complex. In addition, within each group, Best Learned always outperforms Best Traditional, which further demonstrates the superiority of learned methods over traditional methods. We also divide the queries based on the operator type (equality or range) and have the same observation that Best Learned outperforms Best Traditional in both groups.

Among all learned methods, Naru is the most robust and accurate one. It basically has the best q-error across all scenarios and keeps its max q-error within 200. As for query-driven methods, LW-XGB can achieve the smallest q-error in most situations except for max q-error, in which it cannot beat MSCN. We find that the queries which have large errors on LW-XGB and LW-NN usually follow the same pattern: the selectivity on each single predicate is large while the conjunctive of multiple such predicates is very small. This pattern cannot be well captured by the CE features (AVI, MinSel, EBO) adopted LW-XGB/NN. In comparison, MSCN can handle this situation better which may be due to the sample used in its input.

We observe that the same algorithm performs quite differently on different datasets in terms of max q-error. But for the other error metrics like median, the performance is consistent across datasets. This is because max q-error can be easily affected by a few queries. Most methods (e.g., DeepDB, LW-XGB/NN) tend to have bigger max error on larger dataset due to the increasing range of possible cardinality values (number of tuples in total). On the other hand, Naru shows very impressive max q-error on the largest DMV than other datasets. It is because Naru models all columns as discrete values and learns the embedding representation of each value. Since DMV has the smallest domain size (smaller number of discrete values) and also the biggest model budget (larger embedding size), Naru can learn a better representation. MSCN maintains its max error in the same magnitude on all datasets using a random sample, which also leads to the same observation in Sample-A.

### 4.3 What Is the Cost For High Accuracy?

Since learned methods can beat the cardinality estimators used in real database systems by a large margin, can we just directly deploy them? In this section, we examine the cost of these highly accurate

1644

Table 4: Estimation errors on four real-world datasets.

| Estimator | Census | | | | Forest | | | | Power | | | | DMV | | | |
|---|---|---|---|---|---|---|---|---|---|---|---|---|---|---|---|---|
| | 50th | 95th | 99th | Max | 50th | 95th | 99th | Max | 50th | 95th | 99th | Max | 50th | 95th | 99th | Max |
| Traditional Methods | | | | | | | | | | | | | | | | |
| Postgres | 1.40 | 18.6 | 58.0 | 1635 | 1.21 | 17.0 | 71.0 | 9374 | 1.06 | 15.0 | 235 | $2 \cdot 10^5$ | 1.19 | 78.0 | 3255 | $1 \cdot 10^5$ |
| MySQL | 1.40 | 19.2 | 63.0 | 1617 | 1.20 | 48.0 | 262 | 7786 | 1.09 | 26.0 | 2481 | $2 \cdot 10^5$ | 1.40 | 1494 | $3 \cdot 10^4$ | $4 \cdot 10^5$ |
| DBMS-A | 4.16 | 122 | 307 | 2246 | 3.44 | 363 | 1179 | $4 \cdot 10^4$ | 1.06 | 8.08 | 69.2 | $2 \cdot 10^5$ | 1.46 | 23.0 | 185 | $3 \cdot 10^4$ |
| Sample-A | 1.16 | 31.0 | 90.0 | 389 | **1.04** | 17.0 | 67.0 | 416 | **1.01** | 1.22 | 8.00 | 280 | **1.01** | **1.42** | 19.0 | **231** |
| Sample-B | 1.16 | 11.0 | 34.0 | 1889 | **1.04** | 9.83 | 38.0 | 9136 | **1.01** | 1.25 | 8.00 | $2 \cdot 10^5$ | **1.01** | 1.43 | **10.0** | $3 \cdot 10^4$ |
| MHIST | 4.25 | 138 | 384 | 1673 | 3.83 | 66.5 | 288 | $2 \cdot 10^4$ | 4.46 | 184 | 771 | $1 \cdot 10^5$ | 1.58 | 13.8 | 90.8 | $3 \cdot 10^4$ |
| QuickSel | 3.02 | 209 | 955 | 6523 | 1.38 | 15.0 | 142 | 7814 | 3.13 | 248 | $1 \cdot 10^4$ | $4 \cdot 10^5$ | 126 | $1 \cdot 10^5$ | $4 \cdot 10^5$ | $4 \cdot 10^6$ |
| Bayes | **1.12** | **3.50** | **8.00** | 303 | 1.13 | 7.00 | 29.0 | 1218 | 1.03 | 2.40 | 15.0 | $3 \cdot 10^4$ | 1.03 | 1.85 | 12.9 | $1 \cdot 10^5$ |
| KDE-FB | 1.18 | 23.0 | 75.0 | **293** | **1.04** | 5.00 | 17.0 | 165 | **1.01** | 1.25 | 9.00 | **254** | **1.01** | 1.50 | 36.0 | 283 |
| Learned Methods | | | | | | | | | | | | | | | | |
| MSCN | 1.38 | 7.22 | 15.5 | **88.0** | 1.14 | 7.62 | 20.6 | 377 | **1.01** | 2.00 | 9.91 | **199** | 1.02 | 5.30 | 25.0 | 351 |
| LW-XGB | 1.16 | **3.00** | **6.00** | 594 | 1.10 | **3.00** | **7.00** | 220 | 1.02 | 1.72 | 5.04 | 5850 | **1.00** | 1.68 | **6.22** | $3 \cdot 10^4$ |
| LW-NN | 1.17 | **3.00** | **6.00** | 829 | 1.13 | 3.10 | **7.00** | 1370 | 1.06 | 1.88 | 4.89 | $4 \cdot 10^4$ | 1.16 | 3.29 | 22.1 | $3 \cdot 10^4$ |
| Naru | **1.09** | **2.50** | **4.00** | 57.0 | 1.06 | 3.30 | 9.00 | **153** | **1.01** | 1.14 | **1.96** | 161 | **1.01** | **1.09** | **1.35** | **16.0** |
| DeepDB | **1.11** | 4.00 | 8.50 | 59.0 | 1.06 | 5.00 | 14.0 | 1293 | 1.00 | 1.30 | 2.40 | 1568 | 1.02 | 1.86 | 5.88 | 5086 |
| L v.s. T | **win** | **win** | **win** | **win** | lose | **win** | **win** | **win** | **win** | **win** | **win** | **win** | **win** | **win** | **win** | **win** |

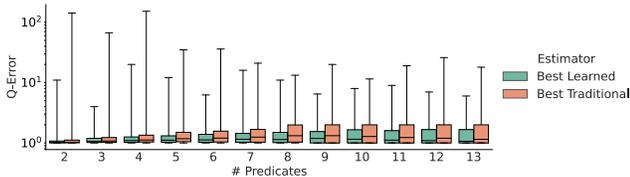

Figure 4: Q-Error comparison between the best learned and traditional method by varying # of predicates on Census.

learned methods. We compare learned methods with database systems in terms of training time and inference time to see whether they can reach the level of DBMS products. Figure 5 shows the comparison result.

**Training Time.** For learned methods, we record the time used to train the models reported in Table 4. For database systems, we record the time to run the statistics collection commands.

Database systems can finish collecting statistics in seconds on all datasets, while learned methods generally need minutes or even hours, which depends on the underlying machine learning model. LW-XGB, which builds on gradient boosted tree, is the fastest learned method. It can be as fast as some DBMS when using fewer trees like in Census and Power dataset. DeepDB is the second fastest, which needs a few minutes to build the SPN model, which also affected by the input sample size and stop conditions. Methods adopt neural networks in general need longer time. Since we use the same epochs on all datasets, Naru's training time highly depends on the data size and platform. With GPU, it only needs 1 minute on Census but takes more than 4 hours on DMV, and this time would be 5× to 15× slower on CPU. GPU acceleration also impacts LW-NN, which takes around 30 minutes to finish training on all datasets but the time can be up to 20× longer on CPU. On the other hand, MSCN exhibits similar training time on the two devices, and GPU is even 3.5× slower than CPU on small datasets. It is because MSCN needs to handle the conditional workflow for minimizing its loss (mean q-error), which becomes slower on GPU and the impact becomes more obvious when the model is smaller.

There is a tradeoff between training time and model accuracy. Neural network methods (Naru, MSCN and LW-NN) trained in an iterative fashion would produce larger error with fewer training iterations. For all these models, we adopt the same epochs reported in the original paper on all datasets, although some models can achieve similar performance with much fewer iterations. For example, using 80% less time, we can train a Naru model on DMV dataset with only slightly performance degrade. However, even if we only run 1 epoch on GPU, it will still be much slower than database systems. We will further explore this trade-off in Section 5.3.

**Inference Time.** We compute the average inference time of the 10K test queries by issuing the queries one by one. Figure 5 shows the result. For database systems, we approximate the time by the latency they return execution plan (without executing it), which should be longer than the real cardinality estimation time due to other overheads such as parsing and binding. Despite of that, all three DBMSs can finish the whole process in 1 or 2 milliseconds. Inference time of learned methods depends on the underlying model. Query-driven methods (MSCN and LW-XGB/NN) are very competitive and can achieve similar or better latency than DBMS (but notice that DBMS's result includes other overheads). It is because they adopt general regression models that directly model the query space and also has been well optimized in terms of implementation. On the other hand, the remaining methods adopt more specialized models and are much slower. SPN model in DeepDB needs around 25ms on three larger datasets and takes an average of 5ms on Census. Naru's inference procedure includes a progressive sampling mechanism, which needs to run the model thousand of times in order to get the accurate result. Its total time is sensitive to the running device, which needs 5ms to 15ms on GPU, and CPU can be up to 20× slower.

The cardinality estimator could be invoked many times during query optimization. Long inference latency can be a blocking issue of bring these accurate learned estimators like Naru and DeepDB into production, especially for OLTP applications with short-running queries. In addition, shortening the inference time of these methods is not a trivial task. Despite the featurization, the bottlenecks of learned methods mostly come from the underlining models, i.e. NN, SPN, XGB, MSCN. To speed up a model's inference time may require techniques, such as model compression/distillation. One exception is Naru, whose bottleneck is, instead of Auto-regressive Model, the dependency of the selectivity computation for each attribute in the progressive sampling procedure, which needs to be done sequentially.

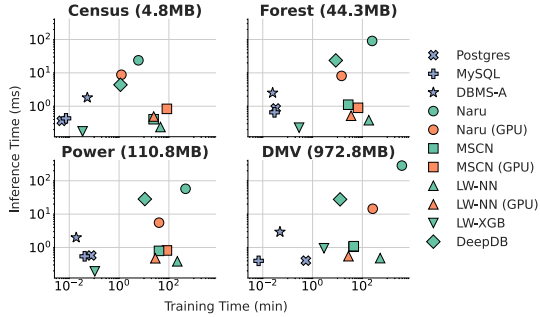

**Figure 5: Training and inference time comparison between learned methods and real database system (MSCN's CPU and GPU results on DMV are overlapped).**

**Hyper-parameter Tuning.** Hyper-parameter tuning is another cost for learned methods. The learned models shown in Table 4 represent the models with the best hyper-parameters. However, without hyper-parameter tuning, learned models may perform very badly. In our experiment, we found the ratio between the largest and the smallest max q-error among models with different hyper-parameters for the same method can be up to $10^5$.

While essential for high accuracy, hyper-parameter tuning is a highly expensive process since it needs to train multiple models in order to find the best hyper-parameters. For example, as shown in Figure 5, Naru spends more than 4 hours in training a single model on DMV with GPU. If five models are trained, then Naru needs to spend 20+ hours (almost a day) on hyper-parameter tuning.

### 4.4 Main Findings

Our main findings of this section are summarized as follows:
- In our experiment, new learned estimators can deliver more accurate prediction than traditional methods in general and among learned methods, Naru shows the most robust performance.
- In terms of training time, new learned methods can be slower than DBMS products in magnitudes except for LW-XGB.
- New learned estimators that based on regression models (MSCN and LW-XGB/NN) can be competitive to database systems in inference time, while methods that model the joint distribution directly (Naru and DeepDB) requires much longer time.
- GPU can speed up the training and inference time of some of the new learned estimators, however it cannot make them as quick as DBMS products and sometimes introduce overhead.
- Hyper-parameter tuning is an extra cost which cannot be ignored for adopting neural network based estimators.

## 5 ARE LEARNED METHODS READY FOR DYNAMIC ENVIRONMENTS?

Data updates in databases occur frequently, leading to a "dynamic" environment for cardinality estimators. In this section, we aim to answer a new question: *Are learned methods ready for dynamic environments?* We want to understand the gap of adopting recent learned methods in real systems. We first discuss how learned methods perform against DBMSs in dynamic environments, then explore the trade-off between the number of updating epochs and accuracy, and finally investigate how much GPU can help.

### 5.1 Setup

**Dynamic Environment.** In a dynamic environment, both model accuracy and updating time matter. Consider a time range $[0, T]$. Suppose that there are $n$ queries uniformly distributed in this time range. Suppose that given a trained initial model, the model update starts at timestamp 0 and finishes at timestamp $t_u$ ($t_u \le T$). For the first $n \cdot \frac{t_u}{T}$ queries, their cardinalities will be estimated using the stale model. For the remaining $n \cdot (1 - \frac{t_u}{T})$ queries, the updated model will be used.

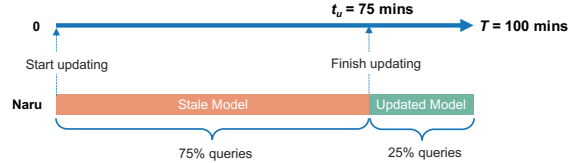

**Figure 6: An illustration of a dynamic environment.**

Figure 6 shows an example. Suppose $T = 100$ mins and Naru spends $t_u = 75$ mins updating its model. Then, Naru needs to estimate the cardinalities for 75% (25%) of the queries using the stale (updated) model. Since many queries will be handled by the (inaccurate) stale model, although Naru performs the best in the static environment, this may not be the case in this dynamic environment.

**Dataset & Workload & Metric** We use the same four real-world datasets as Section 4. We append 20% new data to the original dataset and apply our workload generation method to the updated data to general 10K test queries. That is, the testing workload contains 10K queries. And these queries will be uniformly distributed in $[0, T]$. Here, $T$ is a parameter in our dynamic environment. Intuitively, it represents how "frequent" the data is being updated. For example, if the data is periodically updated every 100 mins, then we can set $T = 100$ mins. We report the 99th percentile q-error of the 10K queries. It is worth noting that we have shown a variety of error metrics (50%, 95%, 99%, and max errors) in Table 4. Based on the results of Table 4, we found that learned methods improve more on the larger errors (99% and max), compared to traditional methods. Since max error is sensitive to outliers, we chose 99% error. To further mitigate the impact of outliers, in our experiment setting, we were using a large number of queries (10,000 queries) for testing. It means that 99% error is the 100th largest error, thus it was not dominated by a few outlier queries.

**Data Update.** We ensure that the appended 20% new data has different correlation characteristics from the original dataset. Otherwise, the stale model may still perform well and there is no need to update the model. To achieve this, we create a copy of the original dataset and sort each column individually in ascending order, which leads to the maximum Spearman's rank correlation between every pair of columns. We randomly pick up 20% of the tuples from this copied dataset and append them to the original dataset.

**Model Update.** The initial models we use are the same as Section 4, which are tuned towards a better accuracy. We follow the original papers of the learned methods to update their models unless stated otherwise. Naru and DeepDB are trained on data. As described in their papers, Naru is updated by one epoch, while DeepDB is updated by inserting a small sample (1%) of the appended data to its tree model. MSCN and LW-XGB/NN use query results as training data. Since the updating procedure is not discussed in the original MSCN paper, we adopt LW-XGB/NN's updating procedure for MSCN. After generating a training workload, we use a sample (5% of the original datasets) to update the query label. LW-XGB and LW-NN originally use 2K and 16K queries for updating accordingly. We assign 10K queries for MSCN as a fair size of training data.

Note that the updating time is different from the training time presented in Figure 5. To update a model quickly, the updating time involves fewer epochs. Also, for query driven methods, they need to add the query results' updating time because this is a major difference between data-driven and query-driven learned methods.

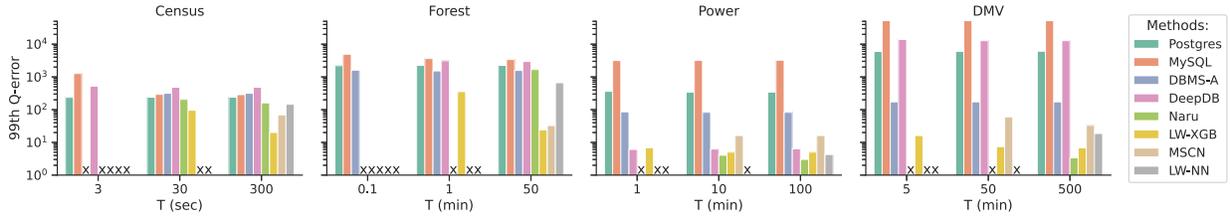

Figure 7: DBMSs *vs* learned methods under different dynamic environments on four datasets.

## 5.2 Which Method Performs the Best in Dynamic Environments?

In this experiment, we test 5 learned methods against 3 DBMSs on CPU. We vary $T$ for each dataset to represent different update frequencies: high, medium, low. Note that our four datasets are different in size, so $T$ is set differently for each dataset. The results are shown in Figure 7. If a model cannot finish updating within $T$, we will put "×" in the figure.

We first compare DBMSs with learned methods. We can see that DBMSs have more stable performance than learned methods by varying $T$. The reason is that DBMSs have very short updating time and almost all the queries are run on their updated statistics. We also observe that many learned methods cannot catch up with fast data updates. Even if they can, they do not always outperform DBMSs. For example, when $T = 50$ mins on DMV, DBMS-A outperforms DeepDB by about 100× since the updated DeepDB model cannot capture correlation change well.

We then compare different learned methods. Overall, LW-XGB can perform better or at least comparable with others in most cases. MSCN and LW-NN do not perform well since they need longer updating time and the stale models process too many queries. DeepDB usually has a very short updating time. However, its updated model cannot capture the correlation change well, thus it does not outperform LW-XGB/NN in most cases. Recall that Naru has a very good accuracy when there is no update. In dynamic environments, however, Naru does not outperform LW-XGB when update frequencies are high or medium. Naru has a similar performance with DBMSs on Census and Forest. This is because Naru uses 1 epoch to update its model, which is not enough to have good accuracy for Census and Forest. For DMV, we have the same observation as [18]. Naru performs well on DMV within 1 epoch. We will discuss this trade-off between updating epochs and accuracy in the next subsection.

In terms of updating time, there is no all-time winner on different datasets. For example, on Census, DeepDB (data driven) is the fastest method, whereas on DMV, LW-XGB (query driven) is the fastest one, although these two methods are the top-2 fastest methods in this experiment. The reason behind this is that the updating time of data driven methods is usually proportional to the size of the data. Intuitively, data driven methods compress the information of the data to the models to represent the joint distribution. When the size of the data gets larger, the complexity of the model should be higher and harder to train. In contrast, query driven methods have the training overhead of generating query labels. However, given a larger dataset and a fixed number of training queries, the complexity of their models do not necessarily become higher. In practice, the choice of using data or query driven methods is really subjective to the applications.

We can observe that each method performs differently on different datasets. One major reason is that in the dynamic environment, there is a trade-off between updating time and estimation accuracy. If a method needs a longer updating time, more queries in the workload will be estimated by the stale model, thus the overall estimation accuracy will degrade.

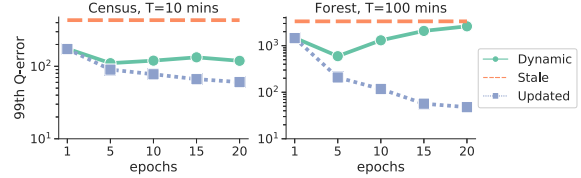

Figure 8: Trade-off (Naru): epochs vs accuracy.

In addition, there are some other reasons that could cause the degradation of accuracy for each individual method. For Naru, one epoch of updating might be insufficient to learn a good updated model. For MSCN, LW-XGB and LW-NN, the ground truth labels are generated from a sample which might introduce errors. For DeepDB, without restructuring the tree, the underlying correlation is assumed unchanged. This assumption might hurt the performance when correlation change happens.

## 5.3 What Is the Trade-off Between Updating Time and Accuracy?

We explore the trade-off between the number of updating epochs and accuracy for learned methods. Due to the space limit, we only show Naru's results on Census and Forest to illustrate this point.

We set $T = 10$ mins on Census and $T = 100$ mins on Forest to ensure Naru with different epochs can finish updating within $T$. Figure 8 shows our results. "Stale" represents the stale model's performance on 10K queries. "Updated" represents the updated model's performance. "Dynamic" represents the Naru's performance (the stale model first and then the updated model) on 10K queries. We can see a clear trade-off of Naru on Forest. That is, "Dynamic" first goes down and then goes up. The reason is that long training time (epochs) makes the model update slow. It leaves more queries executed using the stale mode. Even though more epochs improve the updated model's performance, it hurts the overall performance.

In this Naru experiment, we show the trade-off between updating time and accuracy by varying the number of epochs. There are other ways to achieve this trade-off. For example, for query-driven methods, they need to update the answers to a collection of queries. Using sampling is a nice way to reduce the updating, but it will lead to approximate answers, thus hurting the accuracy. It is an interesting research direction to study how to balance the trade-off for learned methods.

## 5.4 How Much Does GPU Help?

We explore how much GPU can help Naru and LW-NN. We set $T = 100$ mins on Forest and $T = 500$ mins on DMV to ensure they can finish updating within $T$. The results are shown in Figure 9.

We can see that with the help of GPU, LW-NN is improved by around 10× and 2× on Forest and DMV, respectively. There are two reasons for these improvements: (1) LW-NN's training time can be improved by up to 20× with GPU; (2) A well-trained LW-NN (500 epochs) has a good accuracy. For Naru, it is improved by 2× on DMV. However, it does not get improved on Forest. This is because that 1 epoch is not enough for Naru to get a good updated model on Forest, although shorter updating time leaves more queries for the updated model.

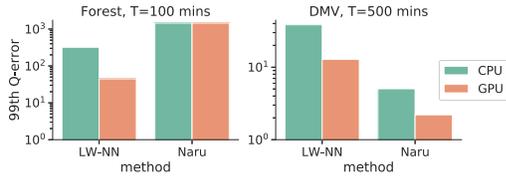

Figure 9: GPU affects the performance.

## 5.5 Main Findings

Our main findings of this section are summarized as follows:
- Learned methods cannot catch up with fast date updates. MSCN, LW-NN, Naru, and DeepDB return large error in dynamic environments for different reasons.
- Within learned methods, there is no clear winner. Naru performs the best when date updates are not frequent, while LW-XGB performs the best in more dynamic environments.
- In terms of updating time, DeepDB is the fastest data-driven method and LW-XGB is the fastest query-driven method.
- There is a trade-off between updating time and accuracy for learned methods. It is not easy to balance the trade-off in practice and requires more research efforts on this topic.
- GPU is able to, but not necessarily, improve the performance. It is important to design a good strategy to handle model updates in order to benefit from GPU.

# 6 WHEN DO LEARNED METHODS GO WRONG?

One advantage of simple traditional methods like histogram and sampling is their transparency. We know that when the assumptions (e.g., attribute-value-independence (AVI), uniform spread) made by these estimators are violated, they tend to produce large q-errors. In comparison, learned estimators are opaque and lack understanding. In this section, we seek to explore scenarios when learned methods do not work well. We run a micro-benchmark to observe how their large error changes when we alter the underlying dataset. We also identify some logical rules that are simple and intuitive but are frequently violated by these learning models.

## 6.1 Setup

**Dataset.** We introduce our synthetic dataset generation procedure. We generate datasets with two columns by varying three key factors: *distribution* (of the first column), *correlation* (between the two columns) and *domain size* (of the two columns). Each dataset contains 1 million rows.

The first column is generated from the genpareto function in scipy [90], which can generate random numbers from evenly distributed to very skewed. We vary the distribution parameter $s$ from 0 to 2, where $s = 0$ represents uniform distribution and the data becomes more skewed as $s$ increases.

The second column is generated based on the first column in order to control the correlation between the two columns. We use $c \in [0, 1]$ to represent how correlated the two columns are. For each row $(v_1, v_2)$, we set $v_2$ to $v_1$ with a probability of $c$ and set $v_2$ to a random value drawn from the domain of the first column with a probability of $1 - c$. Obviously, the two columns are independent when $c = 0$. They are more correlated as $c$ increases and become functional dependent when $c = 1$.

We also consider domain size $d$ (the number of distinct values), which is related to the amount of information contained in a dataset.

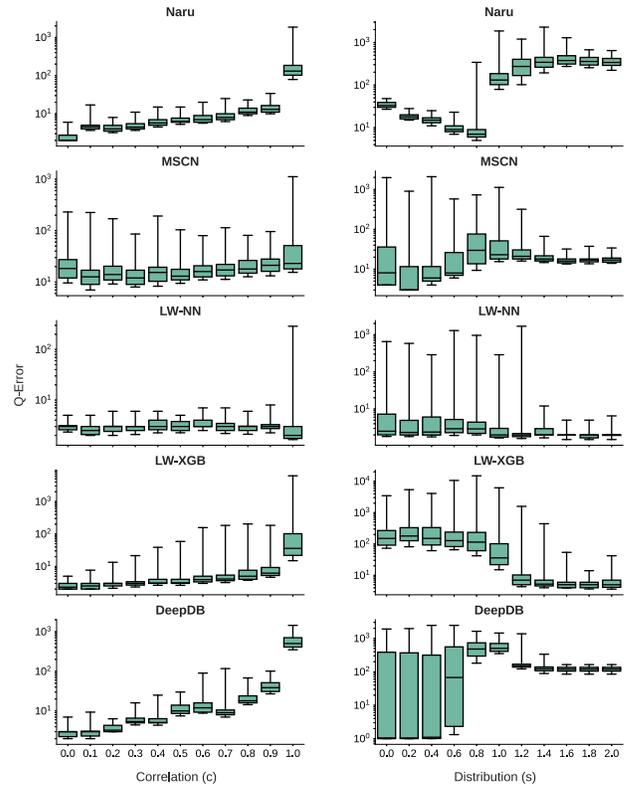

(a) $s = 1.0$, $d = 1000$    (b) $c = 1.0$, $d = 1000$

Figure 10: Top 1% q-error distribution under different correlations (a) and distributions (b).

It can affect the size needed to encode the space for models like Naru. To control the domain size, we convert the generated continuous values into bins. In our experiment, we generate datasets with domain size 10, 100, 1K and 10K.

**Workload.** Since the goal of this experiment is to study the cases when learned methods go wrong, we generate center values from each column's domain independently (OOD) for all the queries in order to explore the whole query space and find as many hard queries as possible. Other workload generation settings are the same as Section 4.

**Hyper-parameter Tuning.** We adopt default hyper-parameters recommended in [30] (RDC threshold = 0.3 and minimum instance slice = 0.01) for DeepDB and fix the tree size of LW-XGB to 128. As for neural network models, we randomly pick up three hyper-parameter settings with 1% size budget using the same way as Section 4 and select one that consistently reports good results. The detailed hyper-parameters used in this experiment can be found in our released code.

## 6.2 When Do Learned Estimators Produce Large Error?

We examine how the accuracy of learned models will be affected by different factors. We train the exact same model on datasets with only one factor varied and the other two fixed, and use the same 10K queries to test the models. Instead of comparing different models, here we aim to observe the performance change for the same model on different datasets. We only exhibit the distribution of the top 1% q-errors to make the trend on large errors more clear. Similar with Section 5, in this experiment, we care more about when learned methods produce large errors.

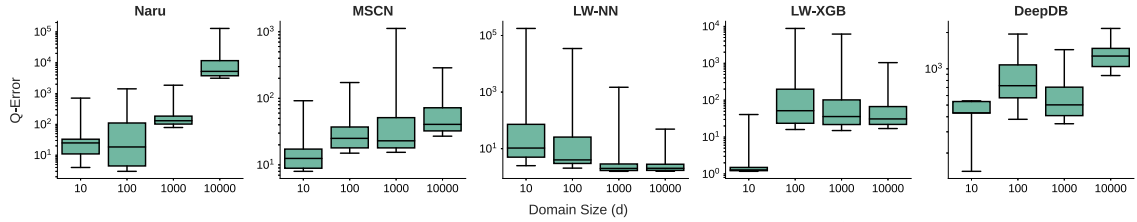
Figure 11: Top 1% error distribution under different domain size ($s = 1.0$, $c = 1.0$).

**Correlation.** A common thing we found when we vary the correlation parameter $c$ is that all methods tend to produce larger q-error on more correlated data. Figure 10a shows the top 1% q-error distribution trend on different correlation degrees with the first column distribution $s = 1.0$ (exponential distribution) and domain size $d = 1000$. It is clear that boxplots in all the figures have a trend to go up when $c$ increases.

Another observation is that the q-error of all estimators rises dramatically (10 ~ 100×) when two columns become functional dependent ($c = 1.0$). This pattern commonly exists on different pairs of $s$ and $d$ values we tested, which indicates that there is space to improve theses learned estimators on highly correlated datasets especially when functional dependency exists.

In our experiment, we found these two observations hold in general on different values for $s$ and $d$.

**Distribution.** Each learned method reacts differently when we change the distribution of the first column, and this reaction also changes when the underlying correlation setting varies. Figure 10b shows the top 1% q-error distribution trend when $s$ goes from 0.0 to 2.0 while fixing the correlation $c = 1.0$ and domain size $d = 1000$. We choose to show the case when two columns are functional dependent because it tends to produce larger errors.

In general, Naru outputs larger max q-errors when data is more skewed ($s > 1.0$), while MSCN, LW-XGB/NN and DeepDB show an opposite pattern. We suspect this difference might be caused by the different basic building blocks used in each method. The common thing shared within the latter approaches is that they all incorporate basic synopsis like sampling or 1D histogram in their models. These statistics might directly record a relatively accurate cardinality for the query involving a frequent value in the dataset, and thus reduce the max error when data is very skewed. If this is true, we can study how to incorporate a similar idea into Naru and make it more robust on skewed data.

Another interesting thing is that unlike max q-error, the 99th percentile q-error (the lower extreme of the boxplot since we only report top 1% q-errors) shows an opposite pattern on MSCN and DeepDB. Here we guess that for both methods, it might be because of the number of queries with very small selectivity increases when $s$ increases. In such cases, the sample feature in MSCN would remain in all zero on many queries, which is not very useful. As for DeepDB, since its leaf node has the AVI assumption, it would produce very large result when the selectivity of each predicate is large but the combined result is very small, which is common when $s$ is large.

**Domain Size.** Figure 11 shows the top 1% q-error distribution on datasets generated under different domain size ($s = 1.0$ and $c = 1.0$). Notice that Naru may use a different model architecture on each domain size to meet the same 1% size budget.

Except for LW-NN, all methods output larger error on larger domain size. Naru exhibits a 100× performance degrade when domain size goes from 1K to 10K. This may be because that the embedding matrix on 10K domain occupies a big portion of the size budget and thus the rest of the model does not have enough capacity to learn the data distribution. Having a more efficient encoding method could mitigate this issue for Naru. LW-XGB shows a very strong result when domain size is 10 and the error becomes 100× bigger on larger domains. MSCN and DeepDB are relatively more robust than other methods but still experience around 10× degrade when domain size increases from 10 to 10K.

It is interesting to see that LW-NN and LW-XGB show opposite trend even though they share the same input feature and optimization goal. It is very likely that this phenomenon is caused by the underlying model they adopt. We suspect that the input query space becomes more "discrete" when the domain size is as small as 10. Therefore a small change in the query predicate can dramatically change the cardinality result or might not affect it at all. It can be hard for the neural network used in LW-NN to learn since compared with the tree-based model in LW-XGB, neural network intuitively fits the data in a more smooth and continuous way.

### 6.3 Do Learned Estimators Behave Predictably?

During our experimental study, we identify some *illogical behaviors* from some of the learned models. For example, when we changed one of the query predicates from [320, 800] to a smaller range [340, 740], the real cardinality decreased, but the estimated cardinality by LW-XGB unexpectedly increased by 60.8%.

This kind of unreasonable behavior caught our attention. The violation of simple logical rules like this could cause troubles for both DBMS developers and users (see Section 6.4 for more discussion). Inspired by the work [83] in the deep learning explanation field, we propose five basic rules for CE. These rules are simple and intuitive which the users may expect cardinality estimators to satisfy:

(1) **Monotonicity:** With a stricter (or looser) predicate, the estimation result should not increase (or decrease).
(2) **Consistency:** The prediction of a query should be equal to the sum of the predictions of queries split from it (e.g., a query with predicate [100, 500] on $A_i$ can be split to two queries with [100, 200] and [200, 500] on $A_i$ respectively and other predicates remain the same).
(3) **Stability:** For any query, the prediction result from the same model should always be the same.
(4) **Fidelity-A:** Result should be 1 for querying on the entire domain (e.g. SELECT * FROM R WHERE $min_i \leq A_i \leq max_i$).
(5) **Fidelity-B:** Result should be 0 for a query with an invalid predicate (e.g. SELECT * FROM R WHERE $100 \leq A_i \leq 10$).

According to these proposed rules, we check each learned estimator and summarize whether it satisfies or violates each rule in Table 5. Some of the rules like Fidelity-B can be fixed with some simple checking mechanisms, however here we only consider the original output of the underlying model used in each estimator in order to see whether these models behave in a logical way natively.

Naru's progressive sampling technique introduces uncertainty to the inference process, which causes the violation of stability. Figure 12 shows an example of the estimation results using Naru to run a query (the actual cardinality is 1036) for 2000 times under this setting. The results are spread over the range of [0, 5992]. This instability also causes Naru to violate monotonicity and consistency rules. The regression-based methods (MSCN, LW-NN, LW-XGB) violate all the rules except for stability. It is not a very surprising result since there is no constraint enforced to the model during both training and inference stages. DeepDB does not violate any rules since it is built on basic histograms and the computation between nodes is restricted to addition and multiplication.

Table 5: Satisfaction and violation of rules by learned estimators. (✓: satisfied, ×: violated)

| Rule | Naru | MSCN | LW-XGB | LW-NN | DeepDB |
|---|---|---|---|---|---|
| Monotonicity | × | × | × | × | ✓ |
| Consistency | × | × | × | × | ✓ |
| Stability | × | ✓ | ✓ | ✓ | ✓ |
| Fidelity-A | ✓ | × | × | × | ✓ |
| Fidelity-B | ✓ | × | × | × | ✓ |

### 6.4 What Will Go Wrong in Production?

We discuss four issues that may appear when deploying (black-box and illogical) learned models in production.

**Debuggability.** It is challenging to debug black-box models like Naru, MSCN and LW-XGB/NN. Firstly, black-box models may fail silently, thus there is a high risk to miss a bug. For example, if there is a bug in the hyper-parameter tuning stage, the model can still be trained and may pass all test cases. Secondly, black-box models make it hard to trace an exception back to the actual bug. If the learned model produces a large error for a given query, it is difficult to tell whether it is a normal bad case or caused by a bug in the code or training data.

**Explainability.** Another issue is that black-box models lack explainability. It brings some challenges for query optimizer version update. We might find a new model architecture improve the accuracy and want to adopt it to the new version. However, it is hard to explain to the database users about which type of query and what kind of scenario will be affected by this upgrade.

**Predicability.** Since learned methods do not follow some basic logic rules, the database system may behave illogically, thus confusing database users. For example, a user would expect a query to run faster by adding more filter conditions. Due to the violation of the monotonicity rule, this may not be the case when the database system adopts a learned model like Naru, MSCN, or LW-XGB/NN.

**Reproducibility.** It is common that a database developer wants to reproduce customers' issues. In order to reproduce the issues, the developer needs information, such as the input query, optimizer configurations, and metadata [80]. However, if the system adopts Naru which violates the stability rule, it would be hard to reproduce the result due to the stochastic inference process.

### 6.5 Main Findings

Our main findings of this section are summarized as follows:
- All new learned estimators tend to output larger error on more correlated data, and the max q-error jumps quite dramatically when two columns are functional dependent.
- Different methods react differently for more skewed data or for data with larger domain size. This might be due to the differences in the choice of models, input features, and loss functions.
- We propose five rules for cardinality estimators and find that all new learned models except for DeepDB violate these rules.
- The non-transparency of the models used in new learned estimators can be troublesome in terms of debuggability, explainability, predicabiltiy, and reproducibility when deployed in production.

## 7 RESEARCH OPPORTUNITY

We have discussed that the high cost (Section 4 and Section 5) and the non-transparency (Section 6) are the two main challenges of applying learned carnality estimators in DBMS. What can we do in order to close these gaps? In this section, we discuss some opportunities in the two research directions.

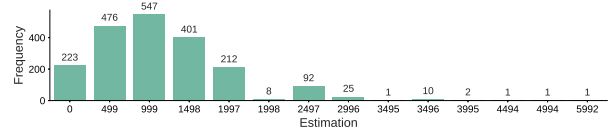

Figure 12: Prediction result of running Naru on the same query 2000 times ($s = 0.0$, $c = 1.0$, $d = 1000$).

### 7.1 Control the Cost of Learned Estimators

**Balance the Efficiency-Accuracy Tradeoff.** Balancing the trade-off between accuracy and training (updating) time as well as inference latency can be an interesting aspect to start with. To retrain a model, simple approximate methods like using a sample instead of full data to calculate the queries' ground-truth or incrementally updating the model, can be leveraged to make neural network models more efficient. Similar ideas in machine learning techniques such as early stop [8] and model compression [11] can also be used to reduce the cost.

**Hyper-parameter Tuning for Learned Estimators.** Hyper-parameter tuning is crucial for new learned models to achieve high accuracy. Algorithms like random search [5], bayesian optimization [78], and bandit-based approaches [46] can be adopted to reduce the cost of obtaining a good hyper-parameter configuration.

Another aspect for hyper-parameter tuning is the goal of tuning. Usually, the goal is to find the configuration with the best accuracy/loss. In the cardinality estimation setting, it is worth doing more exploration to take training/updating time into consideration, because of the trade-off above.

### 7.2 Make Learned Estimators Trustworthy

**Interpret Learned Estimators.** There have been extensive works in machine learning explanation trying to understand why a model makes a specific prediction for a specific input, such as surrogate models [75], saliency maps [79], influence function [35], decision sets [40], rule summaries [76], and general feature attribution methods [52, 83]. These techniques could be leveraged to interpret black box cardinality estimators to some extend. For example, when we get a large error for a query during the test phase, we can use influence function [35] to find the most influential training examples, or we can use shapely value [52] to check the importance of each input feature. However, how effective these methods are in the cardinality estimation setting is still an open problem.

**Handle Illogical Behaviours.** Our study shows that many learned methods do not behave logically. One way to handle this is to define a complete set of logical rules and identify which rules are violated for a certain method. This will add more transparency to each learned method and enable the database developers to know what kind of behavior can be expected from each method. The logical rules we propose in Section 6.3 can be seen as an effort from this perspective. Another way is to enforce logical rules as constraints for model design. There are some existing works in the machine learning community [12, 20, 36]. Similar ideas could be applied to the design of cardinality estimation models.

## 8 MULTI-TABLE SCENARIO

Our paper focuses on understanding single table learned cardinality estimation. In this section, we discuss how to expand current techniques for multi-table cardinality estimation scenario and several challenges involved.

**Extend to Multi-table Scenarios.** MSCN natively supports joins by featurizing the set of base tables, and joins' indicator and predicates. DeepDB supports joins by learning a model on the outer join table if the tables pass the tuple-wise correlation check. LW-XGB/NN and Naru do not support multi-table natively. However, new methods are proposed to support joins based on these two methods. In [94], NeuroCard is proposed as an extension of Naru. They train the auto-regressive model on samples from the full outer join of tables. In [17], they extend LW-XGB/NN to multi-table by training the underlining regression model on queries against a materialized view of the join table.

**Additional Challenges.** In the *training phase*, multi-table methods require to be exposed to the information of the joined tables. This is an additional challenge which will require the model to capture table correlation on more complex data. Moreover, if updates happen in one or more tables in joins, another challenge would be how to update the model accordingly and how to balance the trade-off between update time and model accuracy.

In the *inference phase*, when joins are considered, the number of candidate query plans is exponential to the number of joins. That is, cardinality estimation will be invoked by many times. One way is to adopt the brute force solution, which does inference on every operator in every candidate query plan. But the total inference time could be unacceptably long. Another way is to make inference only on the base tables and calculate the following cardinalities using some formulas. However, this might propagate the errors and hurt the accuracy. Therefore, a big challenge is how to intelligently allocate the inference budget to better balance the accuracy and efficiency trade-off.

## 9 RELATED WORK

**Single Table Cardinality Estimation.** Histogram is the most common cardinality estimation approach and has been studied extensively [1, 6, 21, 25, 26, 31, 47, 57, 60, 61, 71, 73, 74, 81, 86] and adopted in database products. Sampling based methods [22, 48, 77, 93, 97] have the advantage to support more complex predicates than range predicates. Prior work mainly adopts traditional machine learning techniques to estimate cardinality, such as curve-fitting [9], wavelet [58], KDE [29], uniform mixture model [66], and graphical models [14, 24, 88]. Early works [3, 41, 49, 51] also use neural network models to approximate the data distribution in a regression fashion. In comparison, new learned methods have shown more promising results [18, 34].

**Join Cardinality Estimation.** Traditional database systems estimate the cardinality of joins following simple assumptions such as uniformity and independence [42]. Some works [30, 34] can support joins directly, while others [17, 33, 91, 94] study how to extend single table cardinality estimation methods to support join queries. Empirical study [64] evaluates different deep learning architectures and machine learning models on select-project-join workloads. Leis et. al [43] propose an index-based sampling technique which is cheap but effective. Focusing on a small amount of "difficult" queries, some works [70, 92] introduce a re-optimization procedure during inference to "catch" and correct the large errors, while another line of research tries to avoid poor plans by inferring the upper bound of the intermediate join cardinality [7].

**End-to-End Query Optimization.** Recently, more and more works try to tackle the query optimization problem in an end-to-end fashion. Sun et. al [82] propose a learning-based cost estimation framework based on a tree-structured model, which estimate both cost and cardinality simultaneously. Pioneer work [63] shows the possibility of learning state representation of query optimization for the join tree with reinforcement learning, and many follow-up works [38, 55, 87, 96] reveal the effectiveness of using deep reinforcement learning for join order selection. Marcus et. al propose Neo [56], which uses deep learning to generate query plans directly. There are also several end-to-end query optimization systems [4, 80, 99] available in the open-source community.

**Benchmark and Empirical Study in Cardinality Estimation.** Leis et. al [42] propose the Join Order Benchmark (JOB), which is based on the real-world IMDB dataset with synthetic queries having 3 to 16 joins [42]. Unlike JOB, we focus on single table cardinality estimation. Ortiz et. al [64] provide an empirical analysis on the accuracy, space and time trade-off across several deep learning and machine learning model architectures. Our study is different from their work in many aspects. We include both data-driven and query-driven learned methods (whereas they focus on query-driven models) and both static and dynamic settings. Also we try to explore when learned models would go wrong with controlled synthetic datasets and propose simple logical rules to evaluate them. Harmouch et. al [27] conduct an experimental survey on cardinality estimation, but their target is on estimating the number of distinct values, which is different from our paper.

**Machine Learning for Database Systems.** Zhou et. al [100] provide a thorough survey on how ML and DB can benefit each other. In addition to cardinality estimation, ML has the potential to replace and enhance other components in database systems such as indexes [37] and sorting algorithms [39]. Another aspect is to leverage ML to automate database configurations like knob tuning [84, 98], index selection [68], and view materialization [32]. Beyond that, Approximate Query Processing (AQP) [2, 45, 65, 69] engines which support COUNT queries can be potentially adopted for cardinality estimation. Learned AQP [53, 54, 85] is recently in a rising trend. It is interesting to study their effectiveness on supporting cardinality estimation in DBMS.

## 10 CONCLUSION

In our paper, we raised an important but unexplored question: "Are we ready for learned cardinality estimation?". We surveyed seven new learned methods and found that existing experimental studies are inadequate to answer this question. In response, we proposed a unified workload generator and explored whether learned methods are ready for both static environments and dynamic environments, and dived into when learned methods may go wrong. In the end, we identified a number of promising research opportunities.

We concluded that new learned methods are more accurate than traditional methods. However, in order to put them in a well-developed system, there are many missing parts to be resolved, such as low speed in training and inference, hyper-parameter tuning, black-box property, illogical behaviors, and dealing with frequent data updates. As a result, the current learned methods are still not ready to be deployed in a real DBMS. Overall, this is an important and promising direction to be further explored by our community.

## ACKNOWLEDGMENTS

This work was supported in part by Mitacs through an Accelerate Grant, NSERC through a discovery grant and a CRD grant, and WestGrid (www.westgrid.ca) and Compute Canada (www.computecanada.ca). All opinions, findings, conclusions and recommendations in this paper are those of the authors and do not necessarily reflect the views of the funding agencies.